\documentclass[prl,twocolumn,amsmath,amssymb,showpacs]{revtex4-1}
\usepackage{graphicx,psfrag,color,hyperref}
\usepackage{amssymb,latexsym,mathrsfs,amsmath,mathrsfs}
\begin{document}
\def\be{\begin{equation}}
\def\ee{\end{equation}}
\def\bea{\begin{eqnarray}}
\def\eea{\end{eqnarray}}
\def\fr{\frac}
\def\l{\label}

\title{Fluctuating interfaces subject to stochastic resetting}
\author{Shamik Gupta, Satya N. Majumdar, Gr\'egory Schehr}
\affiliation{Laboratoire de Physique Th\'{e}orique et Mod\`{e}les
Statistiques (CNRS UMR 8626), Universit\'{e} Paris-Sud, Orsay, France}
\begin{abstract}
We study one-dimensional fluctuating interfaces of length $L$ where the interface stochastically
resets to a fixed initial profile at a constant rate $r$. For finite
$r$ in the limit $L \to \infty$, the system settles into a nonequilibrium stationary state with
{\em non-Gaussian} interface fluctuations, which we characterize analytically for the Kardar-Parisi-Zhang
 and Edwards-Wilkinson universality class. Our results are corroborated by numerical simulations. We also discuss
the generality of our results for a fluctuating interface in a generic universality class. 
\end{abstract}
\date{\today}
\maketitle

Fluctuating interfaces are paradigmatic nonequilibrium
systems commonly encountered in diverse physical situations,
e.g., propagation of flame fronts in paper sheets, fluid flow in porous
media, vortex lines in disordered superconductors, liquid-crystal
turbulence, and many others. Study of such 
interfaces has many practical applications in the field of molecular
beam epitaxy, crystal growth, fluctuating steps on metals, growing
bacterial colonies or tumor, etc \cite{Barabasi:1995,Halpin:1995,Krug:1997}.
A well-studied model of fluctuating interfaces is the Kardar-Parisi-Zhang (KPZ) equation \cite{Kardar:1986},
which is believed to describe a wide class of such out-of-equilibrium growth processes. 

Earlier studies of the KPZ equation focused on the universal behavior of
the interface roughness, a property which, for instance, in $1+1$
space-time dimensions is characterized by the interface
width $W(L,t)$ at time $t$ for an interface growing over a substrate of linear size $L$. 
It is then known that $W(L,t)$ 
grows algebraically with time as $t^\beta$
for times $t \ll L^z$ where $z$ is the dynamic
exponent, and saturates for
times $t \gg L^z$ to a $L$-dependent value $\sim L^\alpha$. Here,
$\alpha$ is the roughness exponent, while $\beta
= \alpha/z$ is the growth exponent. For the KPZ universality class in
$1+1$ dimensions, one has
$\alpha = 1/2$ and $z = 3/2$ \cite{Barabasi:1995,Halpin:1995,Krug:1997}.
More recently, in this case, significant theoretical progress has shown that in the
growing regime (i.e., for times $t \ll L^z$), the notion of universality
extends beyond the interface width and holds even for the full 
interface height distribution at late times
\cite{Baik:1999,Praehofer:2000,Johansson:2000,Majumdar:2006,exact_KPZ}.
For example, the scaled cumulative distribution of the interface
height fluctuations in a curved
(respectively, flat) geometry is described by the so-called Tracy-Widom (TW) 
distribution
$F_\beta(x)$, with $\beta =2$ (respectively, $\beta = 1$). The
distribution $F_2(x)$
(respectively, $F_1(x)$) characterizes fluctuations at the edge of the spectrum of random
matrices in the Gaussian Unitary Ensemble (GUE) [respectively, Gaussian
Orthogonal Ensemble (GOE)] \cite{Tracy:1996,Tracy:1994}. Height fluctuations
measured in experiments on nematic liquid crystals with both curved and
flat geometry demonstrated a very good agreement with the TW distributions \cite{Takeuchi:2010,Takeuchi:2011}. 

One of the first studied models of interface growth is the so-called Eden model
\cite{Eden:1961}, which aimed at addressing the growth of bacterial
colonies or tumors in mammals. Such growth typically proceeds through stochastic
cell division, and generates an almost compact cell cluster bounded by a
rough interface that within the Eden model has scaling properties 
in the KPZ universality class of interfaces with curved geometry. The growth however 
may be abruptly interrupted with the cell cluster reduced to its initial size by application of
chemicals, as is done, e.g., in chemotherapy to stop the spread of tumor before it
becomes life threatening. It is then interesting to study the effects of
such random interruptions on the growth process. 
In this Letter, we show that random
interruptions, or random resettings, yield novel steady states with
non-Gaussian fluctuations which we characterize analytically.
Here, we mainly focus on the simpler case of flat interfaces, but our
main results can be easily generalized to the case of the curved geometry.  

\begin{figure}[here]
\centering
\includegraphics[width=0.7\linewidth]{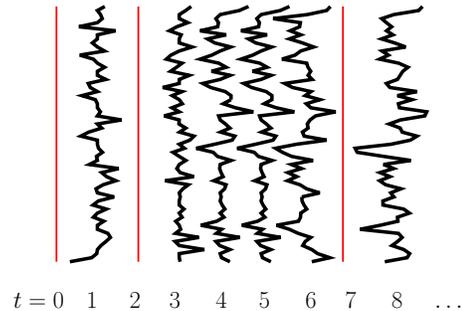}
\caption{ (Color online) Schematic interface evolution with resetting: Starting from a
flat profile, the evolution is interrupted at random times by resetting to the
initial configuration from which it recommences.}
\label{resetting-schematic}
\end{figure}

We consider a $1+1$ dimensional fluctuating interface
characterized by a height field  
$H(x,t)$ at position $x$ and time $t$. Starting from an initially flat profile: $H(x,0) = 0
~\forall~ x$, the heights evolve according to the KPZ equation~\cite{Kardar:1986}:
\be
\fr{\partial H}{\partial t}=\nu \fr{\partial^2 H}{\partial
x^2}+\fr{\lambda}{2}\Big(\fr{\partial H}{\partial x}\Big)^2+\eta(x,t)
\;,
\l{kpz-eom}
\ee
where $\nu$ is the diffusivity, $\lambda$ accounts for the
nonlinearities, while $\eta(x,t)$ is a Gaussian noise of zero mean and correlations $\langle
\eta(x,t)\eta(x',t')\rangle=2D\delta(x-x')\delta(t-t')$. For an interface of length $L$ evolving according to (\ref{kpz-eom}),
the spatially averaged height $\overline{H(x, t)}=\int_0^L {\rm d}x ~H(x,
t)/L$ grows with time with velocity $v_\infty=(\lambda/2)\int_0^L {\rm
d}x ~\langle (\partial H/\partial x)^2 \rangle$. The interface width
$W\equiv W(L,t)$ is defined as the standard deviation of the height about $\overline{H(x,
t)}$. For times larger than a non-universal microscopic timescale
$T_{\rm micro}\sim O(1)$, the width exhibits Family-Vicsek scaling
\cite{Family:1985}: $W(L,t) \sim L^\alpha {\cal W}(t/T^*)$, with 
the crossover timescale $T^* \sim L^z$ corresponding to the scale over which height
fluctuations spreading laterally correlate the entire interface. The
scaling function ${\cal W}(s)$ behaves as a constant as $s\to \infty$,
and as $s^\beta$ as $s \to 0$. At long
times $t \gg T^*$, the dynamics of height fluctuations $h(x,t) \equiv
H(x,t)-\overline{H(x, t)}$ reaches a nonequilibrium
stationary state (NESS) in a {\em finite} system, in which the height distribution
$P_{\rm st}(h)$ is a simple Gaussian \cite{Barabasi:1995}. 

Motivated by the situation where the growth is randomly interrupted,
e.g., by chemicals, as discussed above, 
we study the case where the interface is reset at a fixed rate $r$
to the initial flat configuration. The dynamics with random interruptions, shown schematically in
Fig. \ref{resetting-schematic}, raises a natural question: does it lead to a steady state and if so,  
can one characterize the distribution of the steady state height fluctuations?
Here, we show that indeed random interruptions lead, quite
generically, to a nontrivial steady state {\it even} in the thermodynamic limit
$L\to \infty$, and for a class of 1+1 dimensional models including the KPZ interface,
we compute analytically the height distribution in this steady state. 

Recently, a series of work has shown that resetting dynamics has quite a rich and dramatic 
effect even on a single particle diffusing in
a one-dimensional space $x$ \cite{Evans:2011-1,Evans:2011-2,Evans:2013,Whitehouse:2013}.
The system involves a random walker undergoing diffusion in presence of
resetting, whereby the walker returns to its starting position $x = x_0$
at a constant rate $r$.
The dynamics models the natural search strategy in which a search for 
misplaced belongings after
continuing in vain for a while recommences by returning
to the starting point. 
While in the absence of
resetting, the spatial distribution of the walker is a Gaussian centered
at $x_0$ with width growing diffusively with time as $\sqrt{t}$, a
non-zero $r$ leads to a NESS, with an exponentially decaying profile
centered at $x_0$ \cite{Evans:2011-1,Evans:2011-2,Evans:2013,Whitehouse:2013}. Thus, resetting leads to an otherwise diverging mean
search time finite, thereby increasing the efficiency of the search
strategy. Random walks with restarts have also been extensively used in computer science
as a useful strategy to optimize search algorithms in hard combinatorial problems
\cite{Lovasz:1996,Montanari:2002,Konstas:2009}. 

Our model of resetting of $1+1$ dimensional interface dynamics is a natural extension of the
above mentioned single-particle studies to the case of an extended system comprising many interacting degrees of
freedom.  We note that a recent work also addresses resetting in an
extended system, namely, a one-dimensional coagulation-diffusion process, albeit
with a different resetting strategy \cite{Durang:2013}.

While in the absence of resetting, the dynamics of interface
fluctuations has no steady state in the
thermodynamic limit, we demonstrate here that a non-zero resetting rate
drives the system to a nontrivial NESS, even in the thermodynamic limit.
The NESS obtained is characterized by {\it
non-Gaussian} interface fluctuations, as we demonstrate analytically. In
particular, the stationary
interface width $W_r$ does not scale with the system size, but instead
remains bounded, scaling algebraically with $r$, $W_r \sim r^{-\beta}$,
as $r \to 0$.  We discuss our results for fluctuating interfaces belonging to a generic 
universality class, including the KPZ and the Edwards-Wilkinson (EW)
class. Without resetting, the steady state distribution of height fluctuations
for both the EW and the KPZ class are identical, and are Gaussian. In
contrast, when resetting is switched on, this is not anymore the
case, as resetting carries the information
of the different growth dynamics of the KPZ and the EW class into the 
steady state. To support our
analysis, we present numerical simulations of the interface dynamics. 

We now turn to a derivation of our results. We start with the observation
that for times $t \gg T_{\rm micro}$, when universal scaling behaviors are expected, the 
dynamics 
involves two timescales: (i) $T_{\rm r}
 \sim 1/r$, the average time between two consecutive resets, and (ii) the
crossover time $T^* \sim L^z$. Here we consider
the case $T_{\rm r} \ll T^*$ (but still $T_{\rm r} \gg T_{\rm micro}$),
which is easily achieved in the limit of an infinite system, $L \to
\infty$, with finite $r$. In what follows, we set $L\to \infty$,
or equivalently consider time scales $t \ll T^* = L^z$, such that
the asymptotic dynamics is completely governed by the resetting process.

\begin{figure}[here!]
\includegraphics[width=0.8\linewidth]{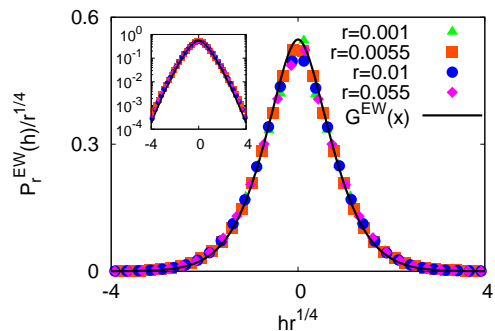}
\caption{ (Color online) EW interface with resetting: Scaling of the distribution of
interface height fluctuations according to Eq. (\ref{ph-scaling}), on
linear (main plot) and log-linear (inset) scales. With $D=\nu=1$, the
points refer to simulation data for $L=2^{14}$, while solid lines for $G^{\rm EW}(x)$ denote analytical
results given by Eq. (\ref{GEW-1}).}
\label{ew-resetting}
\end{figure}

In order to compute the height distribution $P^{\rm
reset}(h,t)$ at time $t$ in presence of resetting, we note that the dynamics in the space of configurations is a Markov process. Indeed,
let us denote by ${\cal C} = \{h(x,t) \}_{0\leq x \leq L}$ a configuration of the whole system of size $L$. The KPZ equation (\ref{kpz-eom})
implies that in the time interval between two successive resetting events, the dynamics of the ``vector'' ${\cal C}$, whose entries are labelled by the space position $x$, is Markovian. The dynamics in configuration space is thus a renewal process. 
Then, at
some fixed time $t$, let the time elapsed since the last
renewal be in the interval $[\tau,\tau+{\rm d}\tau]$, with $0 \le \tau \le t$.
Noting that the probability for this latter event is $re^{-r\tau}{\rm
d}\tau$, we have
\be
P^{\rm reset}({\cal C},t)=\int_0^t {\rm d}\tau
~re^{-r\tau}P({\cal C},\tau)+e^{-rt}P({\cal C},t)\;.
\l{resetting-maineqn-config}
\ee
Here, $P^{\rm reset}({\cal C},t)$ (respectively, $P({\cal C},t)$) is the
probability to be in configuration ${\cal C}$ at time $t$, starting
from an initially flat interface in the presence (respectively, absence) of any resetting. The last term on the right hand side (rhs) of (\ref{resetting-maineqn-config}) accounts for the event when there has not been a single
resetting event in the time interval $[0,t]$. Integrating both sides of Eq.
(\ref{resetting-maineqn-config}) over all the possible configurations ${\cal C}$, 
and noting that $P({\cal C},\tau)$ is normalized to unity for every $\tau$, we check that
$P^{\rm reset}({\cal C},t)$ for every $t$ is also normalized. The
dynamics is Markovian in the configuration space, but is not so for the
relative height $h(x,t)$ at a given point $x$ due to the presence of space derivatives of
the height field on the rhs of (\ref{kpz-eom})
\cite{Bray:2013}.
Nevertheless, Eq. (\ref{resetting-maineqn-config}) being linear, one
obtains the marginal
distribution $P^{\rm reset}(h,t)$ of the height field $h(x,t)$ by
integrating this equation over heights $h(y,t)$ at positions $y \neq x$: 
\be
P^{\rm reset}(h,t)=\int_0^t {\rm d}\tau
~re^{-r\tau}P(h,\tau)+e^{-rt}P(h,t) \;,
\l{resetting-maineqn}
\ee
where $P(h,t)$ is the height distribution in the absence of resetting,
starting from a 
flat initial configuration. 
In the limit $t \to \infty$, we see from (\ref{resetting-maineqn}) that
the system reaches a steady state characterized by the distribution
\be
P_{\rm r}(h)= P^{\rm reset}(h, t \to \infty) = \int_0^\infty {\rm d}\tau~
r e^{-r\tau}P(h,\tau) \;,
\l{ptau-st}
\ee
an exact result valid for any $r$ and $h$. 
Note that due to resetting, a {\it non-local} probability flux
exists only from all $h \ne 0$ values to $h=0$. This leads to a circulation of
probability between a source at $h=0$ and several sinks at $h \ne 0$,
so that the steady state reached is a NESS. 

We consider first the simpler EW equation which corresponds to
$\lambda=0$ in Eq. (\ref{kpz-eom}), thereby leading to an evolution linear in $h$
\cite{Edwards:1982}. In this case, $v_\infty=0$ and
the Family-Vicsek scaling holds with the EW exponents
$\alpha=1/2$ and $z=2$.
Without resetting, the steady state distribution $P_{\rm st}(h)$ at
times $t \gg T^*$ in a finite system is
Gaussian, and is in equilibrium, in contrast to the NESS of a KPZ
interface. For times $t \ll T^*$,
the interface distribution is also Gaussian, but with a non-stationary
width $W^{\rm EW}(t)=D\sqrt{2t/(\pi \nu)}$, for all $t$.  Hence, plugging this Gaussian form for
$P(h,\tau)$ into Eq. (\ref{ptau-st}),
we find that $P^{\rm EW}_{\rm r}(h)$ has the scaling form 
\be
P^{\rm EW}_{\rm r}(h) \sim \sqrt{\gamma} r^{1/4}G^{\rm
EW}(h\sqrt{\gamma}r^{1/4}) \;,
\l{ph-scaling}
\ee
where $\gamma=\sqrt{\pi \nu}/(D2^{3/2})$ and $G^{\rm EW}(x)$ is given by
\be
G^{\rm EW}(x)=\fr{1}{\sqrt{\pi}}\int_0^\infty \fr{{\rm
d}y}{y^{1/4}}\exp\Big(-y-\fr{x^2}{\sqrt{y}}\Big) \;,
\l{GEW-1}
\ee
which is symmetric in $x$, $G^{\rm EW}(-x) = G^{\rm EW}(x)$. From the scaling form in (\ref{ph-scaling}),
one obtains the scaling of the stationary width with $r$ as $W^{\rm EW}_{\rm r} \sim r^{-1/4}$ \cite{footEW}. The integral in (\ref{GEW-1}) can 
be expressed in terms of hypergeometric series. In particular, $G^{\rm
EW}(x)$ behaves asymptotically as
\begin{eqnarray}\label{asympt_GEW}
G^{\rm EW}(x) \sim 
\begin{cases}
&\fr{1}{\sqrt{\pi}}\Big[\Gamma \left(\frac{3}{4}\right)-x^2 \Gamma
\left(\frac{1}{4}\right)+\frac{8}{3} \sqrt{\pi } |x|^3\Big] \;,\; x \to 0 \,, \\
& c|x|\exp[-3/2^{2/3} \, |x|^{4/3}] \;, \; x \to \pm \infty \;,
\end{cases}
\end{eqnarray}
where $\Gamma(x)$ is the Gamma function and $c$ is a computable constant. Interestingly, due to the
$|x|^3$ term in (\ref{asympt_GEW}), $G^{\rm EW}(x)$ is non-analytic close to $x=0$. In the limit $x \to \pm \infty$, the stretched
exponential behavior (\ref{asympt_GEW}) is significantly different from a Gaussian tail. 

\begin{figure}[h!]
\includegraphics[width=0.8\linewidth]{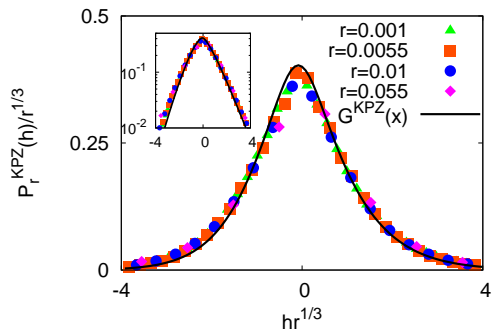}
\caption{ (Color online) KPZ interface with resetting: Scaling of the distribution of
interface height fluctuations according to Eq. (\ref{ph-scaling-kpz}),
on linear (main plot) and log-linear (inset) scales. Here, points refer to
simulation data for $L=2^{15}$, while solid lines for 
$G^{\rm KPZ}(x)$ denote analytical results given by Eq. (\ref{ph-analytic-kpz}).}
\label{kpz-resetting}
\end{figure}

In order to check our prediction
(\ref{ph-scaling}), we now report on results of
numerical simulations performed for a discrete one-dimensional periodic
interface $\{H_i(t)\}_{i=1,2,\ldots,L}$ evolving in times
$t_n = n\Delta t$, with  $n$ an integer and $\Delta t \ll 1$. Starting
from a flat interface, $H_i(0) = 0 ~\forall~ i$, the interface at time step
$t_n$ is reset to its initial configuration with
probability $r\Delta t$, and updated according to the EW
dynamics with probability
$1-r\Delta t$. The results shown in Fig. \ref{ew-resetting} illustrate a
very good agreement with. Evidently, $P^{\rm EW}_{\rm r}(h)$ is highly non-Gaussian. 

We now turn to the KPZ case. Here, it is known that for times $T_{\rm
micro} \ll t \ll T^*$, and for a flat initial profile, the interface height $H(x,t)$ has a deterministic
linear growth with stochastic $t^{1/3}$ fluctuations \cite{exact_KPZ}:
\be
H(x,t)=v_\infty t+(\Gamma t)^{1/3}\chi(x) \;.
\l{H-form}
\ee
Here, $\Gamma\equiv\Gamma(\nu,\lambda,D)$ is a constant, while $\chi$ is a time-independent random variable distributed according to the
TW distribution corresponding to GOE, $f_1(\chi) = F_1'(\chi)$, which can be written explicitly in terms of the Hastings-McLeod solution of the Painlev\'e II equation~\cite{Tracy:1996}. 
In particular, $f_{1}(\chi)$ has
asymmetric non-Gaussian tails \cite{Tracy:1996,Baik:2008}: $f_1(\chi) \approx \exp(-|\chi|^3 /24)$
as $\chi \to -\infty$, while $f_1(\chi) \approx \exp(-2\chi^{3/2} /3)$
as $\chi \to \infty$ \cite{approx-note}.

From Eq. (\ref{H-form}), we get 
\be
h=(\Gamma t)^{1/3}\Big[\chi-\fr{1}{L}\int_0^L {\rm d}x~ \chi(x)\Big] \;.
\ee
Knowing that $f_1(\chi)$ has a finite mean $\langle \chi \rangle < 0$, it follows
from the law of large numbers that in the limit $L \to \infty$, the
second term on the rhs converges to $\langle \chi \rangle$,
so that $\langle h \rangle=0$. In this case, in the limit $\tau \to \infty$, $h \to \infty$, keeping $h/\tau^{1/3}$ fixed, $P(h,\tau)$ takes the scaling form
\be
P(h,\tau) \sim
\fr{1}{(\Gamma \tau)^{1/3}}\widetilde{f}_1\Big(\fr{h}{(\Gamma \tau)^{1/3}}\Big) \;,
\l{kpz-ht-scaling}
\ee
where $\widetilde{f}_1(x)\equiv f_1(x+\langle \chi \rangle)$. From Eq. (\ref{ptau-st}), we see
that in the limit $r \to 0$, $P^{\rm KPZ}_{\rm r}(h)$ has a universal
scaling form, as follows from the fact that the integral (\ref{ptau-st})
in this limit is dominated by
the limit $\tau \to \infty$ where $P(h,\tau)$ can be replaced
by its scaling form~(\ref{kpz-ht-scaling}) for $h \to \infty$, keeping $h/\tau^{1/3}$ fixed. Hence, for $r \to 0$, $h \to
\infty$, keeping $h r^{1/3}$ fixed, we get
\be
P^{\rm KPZ}_{\rm r}(h) \sim (r\Gamma^{-1}) ^{1/3}G^{\rm
KPZ}\left[(r\Gamma^{-1})^{1/3}h\right] \;,
\l{ph-scaling-kpz}
\ee
where the scaling function $G^{\rm KPZ}(x)$ is given by
\be
G^{\rm KPZ}(x)=\int_0^\infty {\rm d}y~\fr{e^{-y}}{y^{1/3}}\widetilde{f}_1\left(\fr{x}{y^{1/3}}\right) \;.
\l{ph-analytic-kpz}
\ee
At variance with $G^{\rm EW}(x)$, the function $G^{\rm KPZ}(x)$ is not symmetric in $x$. On the other hand,
since $\tilde{f}_1$ has zero mean, and has fatter tails as $x \to \infty$ than as
$x \to -\infty$, $G^{\rm KPZ}(x)$ has its maximum
at a negative value $x^*<0$. From (\ref{ph-scaling-kpz}), the stationary
width scales with $r$ as $W^{\rm
KPZ}_{\rm r} \sim r^{-1/3}$. One can check in this case
that $G^{\rm KPZ}(x)$ is analytic close to $x^*$. Its asymptotic
behaviors for $x \to \pm \infty$, obtained from the corresponding
behaviors of $\tilde f_1(x)$ combined with a saddle point analysis, are
\begin{eqnarray}\label{asympt_KPZ}
G^{\rm KPZ}(x) \approx
\begin{cases}
& \exp(-|x|^{3/2}/\sqrt{6}) \;, \; x \to -\infty \\
& \exp(-3^{1/3}x) \;, \; x \to +\infty \;.
\end{cases}
\end{eqnarray}
From (\ref{ph-analytic-kpz}), we see that $G^{\rm KPZ}(x)$ has a
non-analytic behavior as $x \to 0$, where $G^{\rm KPZ}(x) \sim A + B x +
C x^2 \ln x$, with $A,B,C$ being constants. This non-analyticity of the
steady-state distribution at the value to which the system is reset
(here, $x=h=0$, see Fig. \ref{resetting-schematic}) also occurs for EW
interfaces (\ref{asympt_GEW}), as well as in the case of random walks
with resetting \cite{Evans:2011-1,Evans:2011-2}. A comparison
between Eqs. (\ref{asympt_GEW}) and (\ref{asympt_KPZ}) shows that, for $r>0$, the steady state height fluctuations for EW and KPZ dynamics
are quite different. This is in contrast to the case without resetting where they are identical and Gaussian.

To confirm  the scaling form (\ref{ph-scaling-kpz}), we performed numerical simulations of a discrete one-dimensional periodic
interface $\{H_i(t)\}_{1 \le i \le L}$ evolving in discrete times
$t$. The interface is reset to the initial flat configuration with
probability $r$, and updated with probability $1-r$
according to the following dynamics of the ballistic deposition 
model which is in the KPZ universality class~\cite{Barabasi:1995,Halpin:1995,Krug:1997}: 
\bea
H_i(t+1)={\rm max}[H_{i-1}(t),H_i(t)+1,H_{i+1}(t)].
\eea
Comparison between simulations and theory in Fig.
\ref{kpz-resetting} shows a very good agreement. Note that for the comparison, one has to
compute the integral in (\ref{ph-analytic-kpz}) by using the TW GOE distribution whose mean has been shifted to zero, and
then scale the data by a model-dependent fitting
parameter that plays the role of $\Gamma$ in~(\ref{H-form}). We see that as for the EW case,
$P_{\rm r}(h)$ is non-Gaussian.   

For the case of a general interface 
characterized by scaling exponents $\alpha, z$, and $\beta = \alpha/z$,
we now give scaling arguments to compute $P_{\rm r}(h)$. In the limit 
$\tau \to \infty$ and $h \to \infty$, keeping $h /\tau^{\beta}$ fixed, the height
distribution $P(h, \tau)$ quite generally has the scaling form
\be
P(h,\tau) \sim \tau^{-\beta} g\left(h \tau^{-\beta}\right) \;. \l{ph-scaling-general0}
\ee
In this case, the distribution $P_{\rm r}(h)$ is universal in the limit
$r \to 0$ and $h \to \infty$, keeping $hr^\beta$ fixed [see the
discussion following Eq. (\ref{kpz-ht-scaling})]. The associated scaling function is obtained by plugging the behavior in (\ref{ph-scaling-general0}) into Eq. (\ref{ptau-st}), which yields 
\be
P_{\rm r}(h)\sim 
r^{\beta}G(hr^\beta) \;, \; G(x) = \int_0^\infty \frac{dy}{y^\beta} e^{-y} g\left(\frac{x}{y^\beta} \right) \;,
\l{ph-scaling-general}
\ee
implying in particular the behavior of the stationary width 
$W_{\rm r} \sim r^{-\beta}$ as $r \to 0$. The EW and KPZ interfaces
correspond to $\beta = 1/4$ and $\beta = 1/3$ respectively. In the
generic case when $g(x) \sim \exp{(-a x^{\gamma_{\pm}})}$ as $x \to \pm
\infty$, one
obtains by a saddle point analysis of (\ref{ph-scaling-general}) that 
$G(x) \sim \exp{(-b x^{\nu_{\pm}})}$ as $x \to \pm \infty$ with
$\nu_{\pm} = \gamma_{\pm}/(1 + \beta \gamma_{\pm})$. Note also that (\ref{ph-scaling-general}) implies that $G(x)$ is generically
non-analytic at $x=0$.

To conclude, we studied in this work one-dimensional fluctuating
interfaces of length $L$ with the interface stochastically
resetting to a fixed initial profile at a constant rate $r$. For finite
$r$ in the limit $L \to \infty$, the system reaches at long times a 
nonequilibrium stationary state with
non-Gaussian interface fluctuations. We characterized these fluctuations 
analytically for the KPZ
and EW universality class and verified these predictions
via numerical simulations.
For simplicity, we focused here on 
interfaces in a flat geometry, though our results can be easily extended  
to the case of a 
curved geometry. Also, our 
results could possibly be verified in experiments, in particular in the recent ones on liquid crystals \cite{Takeuchi:2010,Takeuchi:2011}.     
 
{\bf Acknowledgements}: We acknowledge support by the ANR grant 
2011-BS04-013-01 WALKMAT and in part by the Indo-French Centre for the 
Promotion of Advanced Research under Project 4604-3.


\begin{thebibliography}{99}


\bibitem{Barabasi:1995}A.-L. Barab\'{a}si and H. E. Stanley, {\em Fractal concepts in surface growth} (Cambridge University Press, 1995).

\bibitem{Halpin:1995} T. Halpin-Healy and Y.C. Zhang, Phys. Rep. {\bf 254}, 215 (1995).

\bibitem{Krug:1997} J. Krug, Adv. Phys. {\bf 46}, 139 (1997). 

\bibitem{Kardar:1986}M. Kardar, G. Parisi, and Y.-C. Zhang, Phys. Rev. Lett. {\bf 56}, 889 (1986).

\bibitem{Baik:1999} J. Baik, P. Deift, and K. Johansson, J. Am. Math. Soc. {\bf 12}, 1119 (1999). 

\bibitem{Praehofer:2000} M. Pr\"ahofer and H. Spohn, Phys. Rev. Lett. {\bf 84}, 4882 (2000).

\bibitem{Johansson:2000} K. Johansson, Comm. Math. Phys. {\bf 209}, 437 (2000).

\bibitem{Majumdar:2006} S. N. Majumdar, in {\it Complex Systems, Volume
LXXXV: Lecture Notes of the Les Houches Summer School}, edited by J.-P.
Bouchaud, M. M\'ezard, and J. Dalibard (Elsevier, 2007) [arXiv:cond-mat/0701193]. 

\bibitem{exact_KPZ} T. Sasamoto and H. Spohn, Phys. Rev. Lett. {\bf 104}, 230602 (2010); T. Sasamoto and H. Spohn, Nucl. Phys. B {\bf 834}, 523 (2010); P. Calabrese, P. Le Doussal, and A. Rosso, Europhys. Lett. {\bf 90}, 20002 (2010); V. Dotsenko, Europhys. Lett. {\bf 90}, 20003 (2010); G. Amir, I. Corwin, and J. Quastel, Comm. Pure and Appl. Math. {\bf 64}, 466 (2011).

\bibitem{Tracy:1996}C. A. Tracy and H. Widom, Comm. Math. Phys. {\bf 177}, 727 (1996).

\bibitem{Tracy:1994}C. A. Tracy and H. Widom, Comm. Math. Phys. {\bf 159}, 151 (1994).

\bibitem{Takeuchi:2010}K. A. Takeuchi and M. Sano, Phys. Rev. Lett. {\bf 104}, 230601 (2010). 

\bibitem{Takeuchi:2011}K. A. Takeuchi, M. Sano, T. Sasamoto, and H. Spohn, Sci. Rep. (Nature) {\bf 1}, 34 (2011). 

\bibitem{Eden:1961}M. Eden, Proc. 4th Berkeley Symp. Math. Stat. Prob. {\bf 4}, 233 (1961).

\bibitem{Family:1985}F. Family and V. Vicsek, J. Phys. A {\bf 18}, L75 (1985).

\bibitem{Evans:2011-1}M. R. Evans and S. N. Majumdar, Phys. Rev. Lett. {\bf 106}, 160601 (2011).

\bibitem{Evans:2011-2}M. R. Evans and S. N. Majumdar, J. Phys. A: Math. Theor. {\bf 44}, 435001 (2011).

\bibitem{Evans:2013}M. R. Evans, S. N. Majumdar, and K. Mallick, J. Phys. A: Math. Theor. {\bf 46}, 185001 (2013).

\bibitem{Whitehouse:2013}J. Whitehouse, M. R. Evans, and S. N. Majumdar, Phys. Rev. E {\bf 87}, 022118 (2013).

\bibitem{Lovasz:1996} L. Lovasz, {\it Random walks on graphs: A survey},
in {\it Combinatronics} (Bolyai Society
for Mathematical Studies, 1996), Vol. 2, p. 1. 

\bibitem{Montanari:2002}A. Montanari and R. Zecchina, Phys. Rev. Lett. {\bf 88}, 178701 (2002). 

\bibitem{Konstas:2009}I. Konstas, V. Stathopoulos, and J.~M. Jose, in
{\it Proc. of the 32nd International ACM SIGIR Conference} (ACM, New
York, 2009), p. 195.

\bibitem{Durang:2013}X. Durang, M. Henkel, and H. Park, arXiv:1309.2107.

\bibitem{Bray:2013}A. J. Bray, S. N. Majumdar, and G. Schehr, Adv. Phys. {\bf 62}, 225 (2013).

\bibitem{Edwards:1982}S. F. Edwards and D. R. Wilkinson, Proc. R. Soc. London Ser. A {\bf 381}, 17 (1982).

\bibitem{footEW}This result is valid here for all $r$, not necessarily
small, since $P(h,\tau)$ is a Gaussian.

\bibitem{Baik:2008}J.~Baik, R.~Buckingham, and J.~DiFranco, Comm. Math. Phys. {\bf 280}, 463 (2008).

\bibitem{approx-note}We use the symbol $\approx$ to mean logarithmic equivalence.

\end{thebibliography}
\end{document}